\documentclass[twocolumn,showpacs,prb,]{revtex4}
\usepackage{amsmath}
\usepackage{graphicx}
\usepackage{dcolumn}
\usepackage{bm}

\begin{document}

\title{Effects of fluctuations and Coulomb interaction on the transition
temperature of granular superconductors. }
\author{I.~S.~Beloborodov$^{1}$, K.~B.~Efetov$^{2,3}$, A.~V.~Lopatin$^{1}$,
and V.~M.~Vinokur$^{1}$}

\address{$^{1}$Materials Science Division, Argonne
National Laboratory, Argonne, Illinois 60439 \\ $^{2}$
Theoretische Physik III, Ruhr-Universit\"{a}t Bochum, 44780 Bochum, Germany \\
$^{3}$ L.~D.~Landau Institute for Theoretical Physics, 117940
Moscow, Russia}

\date{\today}

\begin{abstract}
We investigate the suppression of superconducting transition
temperature in granular metallic systems due to (i) fluctuations
of the order parameter (bosonic mechanism) and (ii) Coulomb
repulsion (fermionic mechanism) assuming large tunneling
conductance between the grains $g_{T}\gg 1$. We find the
correction to the superconducting transition temperature for 3$d$
granular samples and films. We demonstrate that if the critical
temperature $T_c > g_T \delta$, where $\delta$ is the mean level
spacing in a single grain the bosonic mechanism is the dominant
mechanism of the superconductivity suppression, while for critical
temperatures $T_c < g_T\delta$ the suppression of
superconductivity is due to the fermionic mechanism.
\end{abstract}

\maketitle

\section{Introduction}

Being an experimentally accessible electronic system with the
tunable
parameters,~\cite{Valles,experiment,Jaeger,Simon,Beloborodov99,Efetov02}
granular superconductors, offer a unique testing ground for
studying combined effects of disorder, Coulomb interactions and
superconducting fluctuations that govern the physics of disordered
superconductors and are central to mesoscopic physics. One of the
 fundamental questions long calling for investigation is the problem
of suppression of the superconducting critical temperature, $T_c$,
in granular superconductors and the role played in this
suppression by the Coulomb repulsion and superconducting
fluctuations.  In this paper we present a quantitative theory of
the suppression of $T_c$ in granular samples.

The customary belief was that - according to the Anderson
theorem~\cite{Anderson59}- disorder leaves critical temperature of
a superconductor intact.  However this result holds only in the
mean field BCS approximation, and in all the cases where the
extension beyond the BCS approximation is required, one can expect
a noticeable suppression of the critical temperature.

The main mechanisms of the superconductivity suppression are
Coulomb repulsion and superconducting fluctuations.  For example,
disorder shifts significantly the superconducting transition
temperature in the $2d$ thin films
\cite{Ovchinnikov73,Fukuyama81,Finkelstein87,Ishida98,Larkin99}.
The physical reason for the suppression of the critical
temperature is that in thin films the interaction amplitude in the
superconducting channel decreases due to peculiar disorder-induced
interference effects which enhance the effective Coulomb
interaction.  On the technical side, in order to evaluate the
effect of disorder, one should sum a certain class of diagrams
that include, in particular, cooperons and diffusons. In the
subsequent discussion we will be referring to this mechanism of
the superconductivity suppression as to the {\it fermionic}
mechanism.

The superconducting transition temperature can also be reduced by
the fluctuations of the order parameter, the effect being
especially strong in low dimensions.  The corresponding mechanism
of the superconductivity suppression is called the {\it bosonic}
mechanism. In particular, the bosonic mechanism can lead to the
appearance of the insulating state at zero temperature. The
physics of this state can be most easily understood in the case of
a granular sample with weak intergranular coupling: the Cooper
pair can be localized on a single grain if the charging energy is
larger than the Josephson energy corresponding to the
intergranular coupling~\cite{Efetov80}. Later it was
shown~\cite{Fisher90}, that a similar mechanism of Cooper pair
localization appears even in the case of the homogenously
disordered films and the superconductor to insulator transition
was predicted to occur at zero temperature.

In this paper we study the corrections to the superconducting
transition temperature in granular metals perturbativelly. While
this approach is restricted and cannot be used for study of
non-perturbative effects such as the superconductor to insulator
transition, it is useful in a sense that both relevant mechanisms
 of the critical temperature suppression can be studied
systematically within the same framework.  The power of the
perturbative calculation in the study of granular metals was
demonstrated in that it revealed an important energy scale $\Gamma
= g_{\scriptscriptstyle T}\delta$, which was missed for example by
the effective phase functional formalism, where
$g_{\scriptscriptstyle T}$ is the tunneling conductance between
the grains and $\delta $ is the mean energy level spacing for a
single grain, appearing in granular materials. The presence of
this energy scale which has a simple physical interpretation of an
inverse average time that an electron spends in a single grain
before tunneling to one of the neighboring grains~\cite{Efetov},
brings into play new behaviors that are absent in homogeneous
media. In particular, the two {\it different} transport regimes at
high, $T>\Gamma $, and low, $ T < \Gamma $, temperatures
appear.~\cite{Lopatin03} In the high temperature regime the
correction to the conductivity due to the Coulomb interaction
depends logarithmically on temperature in all dimensions, while at
low temperatures the interaction correction to conductivity has
the Altshuler-Aronov form~\cite{Altshuler} and, thus is very
sensitive to the dimensionality of the sample.

In a view of these findings one may expect that the correction to
the superconducting transition temperature can also be different
depending on whether the temperture is larger or smaller than the
energy scale $\Gamma$.

In the present paper we analyze the mechanisms of the suppression
of superconductivity in both temperature regimes. We find that the
fermionic mechanism is temperature dependent and that its
contribution is strongly reduced in the region $T>\Gamma $. In
this regime the bosonic mechanism of the $T_c$ suppression becomes
dominant. In the low temperature regime, $ T<\Gamma $, the
correction to the critical temperature is similar to that obtained
for homogeneously disordered metals. In this regime the fermionic
mechanism plays the major role as long as the intergranular
tunneling conductance is large.

The paper is organized as follows: in Sec.~\ref{summary} we
summarize the results for the suppression of superconducting
transition temperature of granular metals. In
Sec.~\ref{comparison} we compare our results for the suppression
of superconductivity in granular metals with known results for
homogeneously disordered systems. In Sec.~\ref{model} we introduce
the model; the effect of fluctuations and Coulomb interaction on
the superconducting transition temperature is then discussed in
Sec.~\ref{fluctuations}. The mathematical details are relegated to
the Appendixes.

\section{Summary of the results}

\label{summary}

It is convenient to discriminate corrections due to bosonic and
fermionic mechanisms and write the result for the suppression
$\Delta T_{c}$ of the superconductor transition temperature in a
form
\begin{subequations}
\label{sumall}
\begin{equation}
{\frac{{\Delta T_{c}}}{{T_{c}}}}=\left( {\frac{{\Delta
T_{c}}}{{T_{c}}}} \right) _{b}+\left( {\frac{{\Delta
T_{c}}}{{T_{c}}}}\right) _{f}, \label{sum}
\end{equation}
where the two terms in the right hand side correspond to the
bosonic and fermionic mechanisms, respectively. The critical
temperature $T_{c}$ in Eq.~(\ref{sum}) is the BCS critical
temperature.

We find that at high temperatures, $T>\Gamma $, the fermionic
correction to the superconducting transition temperature does not
depend on the dimensionality of the sample
\begin{equation}
\left( {\frac{{\Delta T_{c}}}{{T_{c}}}}\right) _{f}=-c_{1}\,\frac{\delta }{%
T_{c}},\hspace{0.8cm}d=2,3.  \label{totalTc}
\end{equation}
where $c_{1}=7\zeta (3)/2\pi ^{2} - (\ln 2)/4$ is the numerical
coefficient and $d$ is the dimensionality of the array of the
grains.

In the low temperature regime, $T < \Gamma$, the fermionic
mechanism correction to the superconducting transition temperature
depends on the dimensionality of the sample and is given by
\begin{equation}
\left( \frac{\Delta T_{c}}{T_{c}}\right) _{f}=-\left\{
\begin{array}{lr}
\frac{A}{2\pi
\,g_{T}}\ln^{2}\frac{\Gamma}{T_{c}},\hspace{1.6cm}d=3 &
\\
\frac{1}{24\,\pi ^{2}g_{T}}\ln ^{3}\frac{\Gamma }{T_{c}},\hspace{1.3cm}%
d=2 &
\end{array}
\right. ,  \label{mainresult_fermion}
\end{equation}
where $A=g_{T}a^{3}\int d^{3}q/(2\pi )^{3}\varepsilon _{\mathbf{q}%
}^{-1}\approx 0.253$ is the dimensionless constant, $a$ is the
size of a single grain and
\begin{equation}
\varepsilon _{\mathbf{q}}=2g_{T}\sum_{\mathbf{a}}(1-\cos \mathbf{qa})
\label{e0}
\end{equation}
with $\{\mathbf{a}\}$ being the lattice vectors (we consider a
periodic cubic lattice of grains).  Note that in the low
temperature regime $T < \Gamma $ the correction to the critical
temperature in the dimensionality $d=2$ coincides with that
obtained for homogeneously disordered superconducting films upon
the substitution $\Gamma \rightarrow \tau ^{-1}$.

On the contrary, the correction to the transition temperature due
to the bosonic mechanism in Eq.~(\ref{sum}) remains the same in
both regimes and is given by
\begin{equation}
\left( \frac{\Delta T_{c}}{T_{c}}\right) _{b}=-\left\{
\begin{array}{lr}
\frac{14A\zeta (3)}{\pi ^{3}}\;{\frac{1}{{g_{\scriptscriptstyle T}}}},\hspace{1.75cm}d=3 &  \\
{\frac{{ 7 \zeta (3)}}{{2 \pi ^{4}g_{\scriptscriptstyle T}}}}\;
\ln \frac{g_{\scriptscriptstyle T}^{2}\delta }{T_{c}},
\hspace{1.4cm}d=2 &
\end{array}
\right. ,  \label{mainresult_boson}
\end{equation}
where $\zeta(x)$ is the zeta-function and the dimensionless
constant $A$ was defined below Eq.~(\ref{mainresult_fermion}).
Note that the energy scale $\Gamma$ does not appear in this
bosonic part of the suppression of superconducting temperature in
Eq.~(\ref{mainresult_boson}).  This stems from the fact that the
characteristic length scale for the bosonic mechanism is the
coherence length $\xi$ which is much larger than the size of a
single grain. The result for the two dimensional case in
Eq.~(\ref{mainresult_boson}) is written with a logarithmic
accuracy, assuming that $\ln (g_{\scriptscriptstyle T}^{2}\delta
/T_{c})\gg 1$.

The above expression for the correction to the transition
temperature due to the bosonic mechanism was obtained in the
lowest order in the propagator of superconducting fluctuations and
holds therefore as long as the value for the critical temperature
shift given by Eq.~(\ref{mainresult_boson}) is larger than the
Ginzburg region $(\Delta T)_{\scriptscriptstyle G}$
\end{subequations}
\begin{equation}
(\Delta T)_{\scriptscriptstyle G}\sim \left\{
\begin{array}{lr}
\frac{1}{g_{\scriptscriptstyle T}^{2}}
\frac{T_{c}^{2}}{g_{\scriptscriptstyle T}\delta }\hspace{1.4cm}d=3, &  \\
\frac{T_{c}}{g_{\scriptscriptstyle T}}\hspace{2cm}d=2. &
\end{array}
\right.  \label{G_region}
\end{equation}
Comparing the correction to the transition temperature $T_{c}$
given by Eq.~(\ref{mainresult_boson}) with the width of the
Ginzburg region, Eq.~(\ref{G_region}), one concludes that for $3d$
granular metals the perturbative  result~(\ref{mainresult_boson})
holds if
\begin{equation}
T_{c} < g_{\scriptscriptstyle T}^{2}\delta .  \label{intervalt}
\end{equation}
In two dimensions the correction to the transition temperature in
Eq.~(\ref{mainresult_boson}) is only logarithmically larger than
$(\Delta T)_{G}$ in Eq.~(\ref{G_region}). With the logarithmic
accuracy we note that the two dimensional result
(\ref{mainresult_boson}) holds in the same temperature interval
(\ref{intervalt}) as for the three dimensional samples.

Note that inside the Ginzburg region the higher order fluctuation
corrections become important. Moreover, the non perturbative
contributions, in particular, the contributions from
superconducting vortices should be taken into account as well.
These effects destroy the superconducting long range order and
lead to Berezinskii-Kosterlitz-Thouless transition in $2d$
systems.

To summarize our results, we find that the correction to the
superconducting transition temperature of granular metals comes
from two different mechanisms; the dominant mechanism depends on
temperature range.
\newline
(i) In the low temperature regime, $T<\Gamma $, the fermionic
mechanism is the main mechanism of the suppression of $T_{c}$ and
the correction to the transition temperature is given by
Eq.~(\ref{mainresult_fermion}).
\newline (ii) In the high
temperature regime, $T>\Gamma $, the dominant mechanism is
bosonic.  At moderate temperatures, $T<g_{\scriptscriptstyle
T}^{2}\delta $, the correction to the transition temperature is
perturbative and is given by Eq.~(\ref{mainresult_boson}), while
at higher temperatures $T>g_{\scriptscriptstyle T}^{2}\delta $ the
superconducting transition temperature must be determined by
considering the critical fluctuations in the effective
Ginzburg-Landau functional.

\section{Comparison of the suppression of superconductivity in
granular and homogeneously disordered systems}

\label{comparison}

In this section we compare our results for the suppression of
superconductivity with the known results obtained for homogenously
disordered superconductors. We begin our discussion with
homogeneously disordered samples briefly reminding what is known
about suppression of superconductivity in this case.

Both mechanisms of the suppression of superconductivity in
homogeneously disordered films were discussed in several
publications~\cite{Ovchinnikov73,Fukuyama81,Finkelstein87,Fisher90,Ishida98}.
In particular, for films with thickness $d$ such that $l \ll d \ll
\xi$ where $l$ is the electron mean free path and $\xi$ is the
coherence length it was shown that the result for the suppression
of superconducting critical temperature can be written in
analogous form with Eq.~(\ref{sum}), $\Delta T_c/T_c = (\Delta
T_c/T_c)_f + (\Delta T_c/T_c)_b$, where~\cite{Ovchinnikov73}
\begin{subequations}
\label{homogeneous}
\begin{equation}
\label{5a} \left( \frac{\Delta T_{c}}{T_{c}}\right) _{f} = -
\frac{1}{24\,\pi ^{2}g}\ln ^{3}[1/(\tau T_{c})],
\end{equation}
and
\begin{equation}
\left( \frac{\Delta T_{c}}{T_{c}}\right) _{b} = -
\frac{7\zeta(3)}{2\pi^4 g}\ln [g/(\tau T_{c})].
\end{equation}
\end{subequations}
Here $g \gg 1$ is the film conductance (per one spin component)
and $\tau$ is the elastic electron mean free time. One can see
from Eqs.~(\ref{homogeneous}) that in the regime of large
conductance within the logarithmic accuracy the fermionic
mechanism is the dominant one. At the same time, if the
conductance is not too large both corrections become of the order
of one and the bosonic mechanism becomes very important as well.
In this regime the suppression of superconductivity should be
considered non-perturbativelly as in Ref.~\onlinecite{Fisher90}
for the bosonic mechanism and in Ref.~\onlinecite{Finkelstein87}
for the fermionic mechanism.

In granular superconductors situation is different due to
appearance of the energy scale $\Gamma = g_{\scriptscriptstyle
T}\delta$.
As one can see from Eqs.~(\ref{sumall}) both mechanisms of the
suppression of superconductivity are important. In the limit of
high temperatures $T > \Gamma $ the interference effects in
granular metals are suppressed and that is why the fermionic
mechanism is strongly reduced, Eq.~(\ref{totalTc}). The shift of
the superconducting critical temperature in this region is defined
by the bosonic mechanism and has a classic origin. In the low
temperature limit $T < \Gamma$ quantum interference effects become
important therefore the suppression of superconductivity is
defined by the fermionic mechanism. The fact that in the low
temperature regime the correction to the superconducting
transition temperature for a granular samples can be obtained from
the corresponding result for the homogenously disordered samples
via the substitution of the effective diffusion coefficient $ D =
g_T \delta a^2$,  suggests that the universal low temperature
description proposed in Ref.~\onlinecite{Universal} can be
generalized to include the superconducting channel.

\section{The model}

\label{model}

Now we turn to the quantitative description of our model and
derivation of Eqs.~(\ref{sum}, \ref{mainresult_fermion},
\ref{mainresult_boson}). We consider a $d-$dimensional array of
superconducting grains in the metallic state. The motion of
electrons inside the grains is diffusive and they can tunnel
between grains. We assume that if the Coulomb interaction were
absent, the sample would have been a good metal at $T>T_{c}$.

The Hamiltonian of the system of the coupled superconducting
grains is:

\begin{subequations}
\label{hamiltonian}
\begin{equation}
\hat{H}=\hat{H}_{0}+\hat{H}_{c}+\hat{H}_{t}.  \label{hamiltonian1}
\end{equation}%
The term $\hat{H}_{0}$ in Eq.~(\ref{hamiltonian1}) describes isolated
disordered grains with an electron-phonon interaction
\begin{equation}
\hat{H}_{0}=\sum\limits_{i,k}\epsilon _{i,k}a_{i,k}^{\dagger
}a_{i,k}-\lambda \sum\limits_{i,k,k^{\prime }}a_{i,k}^{\dagger
}a_{i,-k}^{\dagger }a_{i,-k^{\prime }}a_{i,k^{\prime }}+\hat{H}_{imp},
\label{e5}
\end{equation}%
where $i$ labels the grains, $k\equiv (\mathbf{k},\uparrow )$,
$-k\equiv (-\mathbf{k},\downarrow )$; $\lambda >0$ is the
interaction constant; $a_{i,k}^{\dagger }(a_{i,k})$ are the
creation (annihilation)
operators for an electron in the state $k$ of the $i$-th grain, and $\hat{H}%
_{imp}$ represents the elastic interaction of the electrons with
impurities. The term $\hat{H}_{c}$ in Eq.~(\ref{hamiltonian1})
describes the Coulomb repulsion both inside and between the grains
and is given by
\begin{equation}
\hat{H}_{c}={\frac{{\ e^{2}}}{{\
2}}}\,\sum_{ij}\,\hat{n}_{i}\,C_{ij}^{-1}\, \hat{n}_{j},
\label{Hc}
\end{equation}
where $C_{ij}$ is the capacitance matrix and $\hat{n}_{i}$ is the
operator of the electron number in the $i$-th grain.

Eq. (\ref{Hc}) describes the long range part of the Coulomb interaction,
which is simply the charging energy of the grains. The last term in the
right hand side of Eq.~(\ref{hamiltonian1}) is the tunneling Hamiltonian
\begin{equation}
\hat{H}_{t}=\sum_{ij,p,q}t_{ij}a_{i,p}^{\dagger }a_{j,q},  \label{Ht}
\end{equation}
where $t_{ij}$ is the tunneling matrix element corresponding to
the points of contact of $i$-th and $j-$th grains and $p$, $q$
stand for the states in the grains.

In the following section we will study effects of fluctuations on
the superconducting transition temperature of granular metals
based on the model defined by Eqs.~(\ref{hamiltonian}).

\section{Effects of fluctuations and Coulomb interaction on transition
temperature} \label{fluctuations}

The superconducting transition temperature can be found by
considering corrections to the anomalous Green function $F$ due to
fluctuations of the order parameter and Coulomb interaction in the
presence of infinitesimal source of pairs
$\Delta$.~\cite{Ovchinnikov73} Without account of fluctuations and
interaction effects, the anomalous Green function $F$ is given by
the expression~\cite{AGD}
\end{subequations}
\begin{equation}
F(\xi ,\varepsilon _{n})=\Delta /(\varepsilon _{n}^{2}+\xi ^{2}),
\end{equation}
where $\xi =\mathbf{p}^{2}/2m-\mu ,$ and $\varepsilon _{n}=2\pi T(n+1/2)$ is
the fermionic Matsubara frequency. The suppression of the transition
temperature $T_{c}$ is determined by the correction to the function $F(\xi
,\varepsilon _{n})$
\begin{equation}
\frac{\Delta T_{c}}{T_{c}}={\frac{T}{\Delta }}\,\int d\xi
\,\sum\limits_{\varepsilon _{n}}\delta F(\xi ,\varepsilon _{n}),
\label{ovchin}
\end{equation}%
where the function $\delta F(\xi ,\varepsilon _{n})$ represents
the leading order corrections to the anomalous Green function
$F(\xi ,\varepsilon _{n})$ due to pair density fluctuations and
Coulomb interaction.
\begin{figure}[t]
\label{beforeaverage} \resizebox{.43\textwidth}{!}{%
\includegraphics{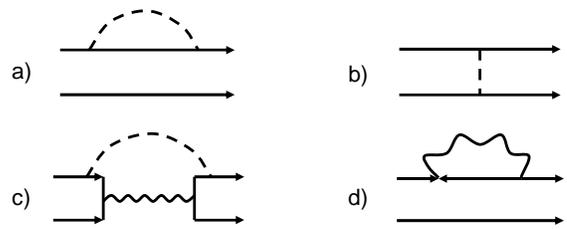}} \vspace{0.6cm}
\caption{Diagrams a) - c) describe the correction to the superconducting
transition temperature due to Coulomb repulsion. The diagram d) describes
correction to the transition temperature due to superconducting
fluctuations. All diagrams are shown before averaging over the impurities.
The solid lines denote the electron propagators, the dashed lines describe
screened Coulomb interaction and the wavy lines describe the propagator of
superconducting fluctuations.}
\end{figure}
The function $\delta F(\xi ,\varepsilon _{n})$ can be found by
means of two different methods which lead to identical results:
(i) solving the Usadel equation with the help of perturbation
theory in powers of the fluctuating order parameter and potential
and further averaging over them using the Gaussian
approximation~\cite{Ovchinnikov73} or (ii) using the diagrammatic
technique. For our purpose we choose the diagrammatic approach.
All diagrams (before impurity averaging) which contribute to the
suppression of the transition temperature in Eq.~\ref{ovchin} are
shown in Fig.~1. One can see that there exist two qualitatively
different classes of diagrams.  First, the diagrams a) - c)
describe corrections to the transition temperature due to Coulomb
repulsion and represent the so called fermionic mechanism of the
suppression of superconductivity. The second type, diagram d),
describes a correction to the transition temperature due to
superconducting fluctuations and represents the bosonic mechanism.
It may seem surprising that we classify the diagram (c) as
belonging to the fermionic mechanism, since this diagram contains
both Coulomb and Cooper pair propagators. The reason is that, as
we will show below (see also
Ref.~\onlinecite{Finkelstain_review}), there are dramatic
cancellations between contributions of diagrams of the types (a,b)
and (c). It is this cancellation that is responsible for the
smallness of the contribution of the fermionic mechanism at high
temperatures $T>\Gamma $. The diagrams of type (c) were not taken
into account in Ref.~\onlinecite{BLV_2004}, where a different
result for the suppression of the transition temperature was
obtained~\cite{Aleiner}. In what follows we consider both
mechanisms of the suppression of superconductivity in details.

\subsection{Suppression of superconductivity due to fluctuations of the
order parameter: bosonic mechanism}

\begin{figure}[t]
\resizebox{.48\textwidth}{!}{\includegraphics{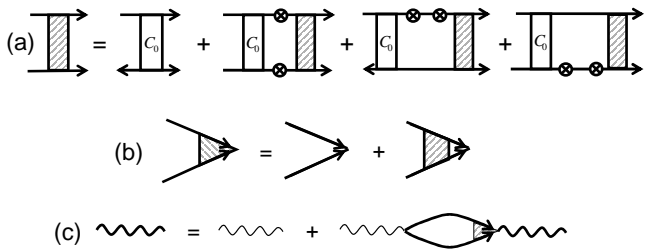}}
\vspace{0.6cm} \caption{Diagrams (a) define the Cooperon
propagator (shaded rectangle), Eq.~(\ref{cooperon1}), in terms of
the single grain Cooperon $C_0.$ Diagrams (b) describe the
renormalization of the BCS interaction vertex due to impurities.
The superconducting propagator, Eq.~(\ref{K1}), is represented by
the thick wavy line and is defined by the diagrams (c) where the
thin wavy line denotes the bare superconducting propagator.  The
solid lines denote the propagator of electrons and the tunneling
vertices are denoted as the circles.} \label{CooperonFig}
\end{figure}

In this section we consider the suppression of the superconducting
transition temperature in granular metals due to fluctuations of
the order parameter (bosonic mechanism). We will use the
diagrammatic technique developed in
Refs.~\onlinecite{Beloborodov99,Efetov}. The main building block
of the diagrams that will be considered in this section is the
Cooperon propagator defined by the diagrams shown in
Fig.~\ref{CooperonFig}~a.  In the regime under consideration all
characteristic energies are much less than the Thouless energy
$E_{\scriptscriptstyle T}=D/a^{2},$ where $D$ is the diffusion
coefficient. This allows us to use the zero dimensional
approximation for a single grain Cooperon  propagator
$C_0^{-1}=\tau| \Omega_n| $. The resulting expression for the
Cooperon is
\begin{equation}
C(\Omega _{n},\mathbf{q})=\tau ^{-1}(|\Omega _{n}|+\varepsilon _{\mathbf{q}%
}\delta )^{-1},  \label{cooperon1}
\end{equation}
where $\mathbf{q}$ is the quasi-momentum and $\Omega _{n}$ is the
bosonic Matsubara frequency. The parameter $\varepsilon
_{\mathbf{q}}$ in the right hand side of Eq.~(\ref{cooperon1})
appears due to the electron tunneling from grain to grain, it was
defined in Eq.~(\ref{e0}).

The propagator of superconducting fluctuations,
$K(\Omega_n\mathbf{q})$ is defined by the diagrams shown in
Fig.~\ref{CooperonFig}~b and~c. They result in the following
expression \label{bosonic1}
\begin{equation}
K(\Omega_n ,\mathbf{q}) = \left[\ln\frac{T}{T_c} +
\psi\left(\frac{1}{2} + \frac{|\Omega_n| +
\varepsilon_{\mathbf{q}}\delta}{4 \pi T}\right) -
\psi\left(\frac{1}{2}\right) \right]^{-1},  \label{K1}
\end{equation}
with $\psi(x)$ being the digamma function.
\begin{figure}
\resizebox{.43\textwidth}{!}{\includegraphics{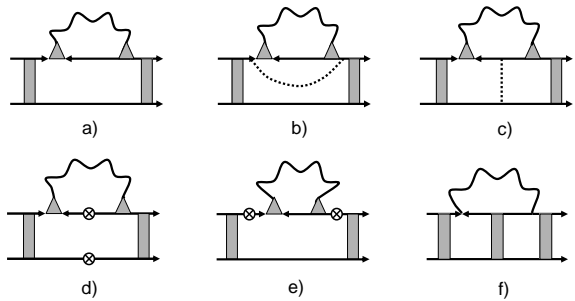}}
\vspace{0.6cm} \caption{Diagrams describing correction to the
transition temperature due to superconducting fluctuations
(bosonic mechanism). The diagrams were obtained after averaging
the diagram d) in Fig.~1 over the disorder. The solid lines denote
the propagator of electrons, the dotted lines describe the elastic
interaction of electrons with impurities and the wavy lines
describe the propagator of superconducting fluctuations. The
shaded rectangle and triangle denote the Cooperon, see
Eq.~(\ref{cooperon1}), and impurity vertex of granular metals
respectively. The tunneling vertices are denoted as the circles. }
\label{BososnicFig}
\end{figure}

The diagrams describing the correction to the transition
temperature in the lowest order with respect to the
superconducting fluctuation propagator $K(\Omega_n ,\mathbf{q})$
are shown in Fig.~\ref{BososnicFig}.

\begin{figure}[t]
\includegraphics[width=.65\linewidth]{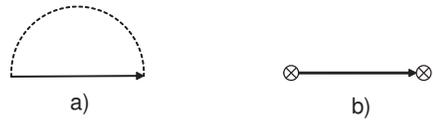}
\caption{Diagrams describing self-energy corrections to the single
electron propagator due to a) elastic interaction of electrons
with impurities and b) electron hopping. The solid lines denote
the bare propagator of electrons and the doted line describes the
elastic interaction of electrons with impurities. The tunneling
vertices are described by the circles.} \label{fig:2}
\label{Self_Energy}
\end{figure}

Deriving the analytical result for the diagrams in
Fig.~\ref{BososnicFig} it is important to take into account the
fact that the single electron propagator itself gets renormalized
due to electron hopping. Tunneling processes give rise to an
additional term in the self-energy part of the single electron
propagator, see Fig.~\ref{Self_Energy}
\begin{equation}
\tau ^{-1} = \tau _{0}^{-1} + 2dg_{T}\delta ,  \label{self}
\end{equation}
where $\tau _{0}$ is the unrenormalized electron mean free time. Although
the second term in the right hand side of Eq.~(\ref{self}) is much smaller
than the first one, it is important to keep it because the leading order
contribution in $\tau _{0}^{-1}$ to the correction to superconducting
transition temperature cancels.

The contribution of each diagram in Fig.~\ref{BososnicFig} to the
suppression of superconducting critical temperature is presented
in Appendix~A. Here we write the final expression for the
contribution of the diagrams (a)-(f) in Fig.~\ref{BososnicFig} to
the superconducting transition temperature
\begin{widetext}
\begin{equation}
\left( \frac{\Delta T_c}{T_c}\right)_b = - \pi T^2 \delta
\sum\limits_{{\bf q}} \left[
\sum\limits_{\varepsilon_n(\varepsilon_n -\Omega_n) > 0}
\frac{K(\Omega_n, {\bf q})[2|\varepsilon_n| + |2\varepsilon_n -
\Omega_n| + \varepsilon_{\bf
q}\delta]}{\varepsilon_n^2(|2\varepsilon_n - \Omega_n| +
\varepsilon_{\bf q}\delta)^2 }
 - \sum\limits_{\varepsilon_n(\varepsilon_n -\Omega_n)<
0}\frac{K(\Omega_n, {\bf q})}{\varepsilon_n^2(|\Omega_n| +
\varepsilon_{\bf q}\delta)} \right]. \hspace{1cm} \label{K}
\end{equation}
\end{widetext}
Here the summation is going over the quasi-momentum, $\mathbf{q}$
and over the fermionic, $\varepsilon_n = \pi T(2n+1)$ and bosonic,
$\Omega_n = 2\pi T n$ Matsubara frequencies.

The main contribution to the suppression of superconducting
transition temperature in Eq.~(\ref{K}) comes from the region of
classical fluctuations and is given by the term with $\Omega
_{n}=0$. Performing summation over the fermionic Matsubara
frequency $\varepsilon _{n}$ in Eq.~(\ref{K}) we obtain the
following result
\begin{equation}
\left( \frac{\Delta T_{c}}{T_{c}}\right) _{b}=-\frac{14\zeta
(3)}{\pi ^{3}} \sum\limits_{\mathbf{q}}\frac{1}{\varepsilon
_{\mathbf{q}}},  \label{b}
\end{equation}
where $\zeta (x)$ is the zeta-function. Equation~(\ref{b}) for the
suppression of superconducting transition temperature is valid
outside the Ginzburg region otherwise the lowest order
approximation in the superconducting propagator which we used to
derive Eq.~(\ref{b}) is not justified. Performing summation over
the quasi-momentum $\mathbf{q}$ in Eq.~(\ref{b}) we obtain the
final result for the suppression of superconductivity due to the
bosonic mechanism, Eq.~(\ref{mainresult_boson} ). The singularity
in two dimensional case should be cut at momenta $\mathbf{q}
_{min}^{2}=a^{-2}\,(T_{c}/g_{T}^{2}\delta )$ in accordance with
the expression for the Ginzburg number
\begin{equation}
Gi\sim \left\{
\begin{array}{lr}
\frac{1}{g_{T}}\hspace{1.5cm}2d, &  \\
\frac{1}{g_{T}^{2}}\frac{T_{c}}{g_{T}\delta }\hspace{1cm}3d. &
\end{array}
\right.  \label{Ginzburg}
\end{equation}
The divergence of the correction to the transition temperature in $%
2d$, Eq.~(\ref{b}), means that flucutuations destroy the
superconducting long range order, which is to be recovered by
introducing the artificial cutoff ${\bf q}_{\min }$. Then the
critical temperature which we calculate should be viewed as a
crossover temperature rather than the temperature of a true phase
transition. However, since experimentally such a temperature marks
a sharp decay of the resistivity, the notion of the transition
temperature still makes a perfect sense.

It follows from Eq.~(\ref{Ginzburg}) that for $3d$ granular metals
the Ginzburg number is small in comparison with the right hand
side of Eq.~(\ref{b}) and the Gaussian approximation for bosonic
mechanism is justified for temperatures
$T_{c}<g_{\scriptscriptstyle T}^{2}\delta $. In $2d$ case the
result (\ref{b}) holds with the logarithmic accuracy in the same
temperature interval $g_{\scriptscriptstyle T}^{2}\delta >> T_c$.

The correction to the transition temperature, Eq.~(\ref{b}), can
be interpreted as a contribution of the fluctuations of the
superconducting order parameter. These fluctuations can be
considered as a virtual creation of Cooper pairs which are bosons.
That is why we call this mechanism of the suppression of the
superconductivity bosonic. The correction $\Delta T_{c}$, can be
rather easily obtained via the Ginzburg-Landau expansion in the
order parameter $\Delta \left( \mathbf{r}\right) $ near the
critical temperature. For the granular system the free energy
functional $F\left[ \Delta \right] $ can be written as follows
\begin{subequations}
\begin{equation}
F = \sum_{i}\left[ \frac{\tilde \tau}{\delta} \left| \Delta
_{i}\right| ^{2}+\frac{b}{2}\left| \Delta _{i}\right| ^{4}\right]
+\sum_{i,j}J_{ij}\left| \Delta _{i}-\Delta _{j}\right| ^{2},
\label{e1}
\end{equation}
where the coefficients $\tilde \tau$ and $b$ are given by
\begin{equation}
\tilde \tau = \frac{T - T_c}{T_c}, \hspace{0.3cm} b = {{ 7\zeta(3)
 }\over { 8\, \pi^2 T_c^2 \delta}},
\end{equation}
\end{subequations}
with $J_{i,i \pm 1} = J_0 = \pi g_T/ 16\, T_c$ being the Josephson
coupling between the grains. Neglecting the quartic term in
Eq.~(\ref{e1}) we obtain the propagator of $\left\langle \Delta
_{\bf q}\Delta _{-\bf q}\right\rangle _{0}$ in the form
\begin{equation}
\left\langle \Delta _{\bf q}\Delta _{-\bf q}\right\rangle
_{0}=\frac{ T_c\delta }{ \tilde \tau + (\pi \delta /8 T_c) \,
\varepsilon _{\mathbf{q}}}. \label{e2}
\end{equation}
The correction to the transition temperature can be found by
calculating the first order in $b$ contribution to the self energy
$\Sigma ^{\left( 1\right) }$:
\begin{equation}
\Sigma ^{\left( 1\right) }=  2 b \sum_{\bf q} {\delta \over
{\tilde \tau + (\pi \delta /8 T_c) \varepsilon _{\mathbf{q}} }}.
\label{e3}
\end{equation}
This correction renormalizes the critical temperature. Putting
$\tilde \tau =0$ in Eq. (\ref{e3}) we come to the result expressed
by Eq. (\ref{b}).

The fermionic mechanism of the suppression of the superconductivity is more
complicated and we consider it in the next section.

\subsection{Suppression of superconductivity due to Coulomb repulsion:
fermionic mechanism}

\begin{figure}[t]
\includegraphics[width=.95\linewidth]{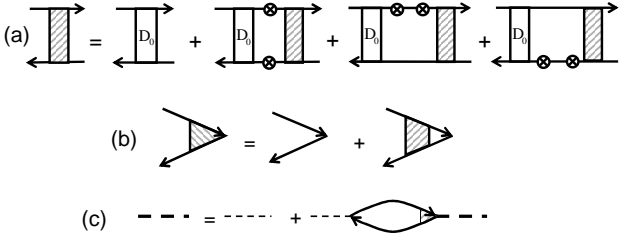}
\caption{Diagrams (a) define the Diffusion propagator $D(\Omega_n,
{\bf q})$, Eq.~(\ref{diffuson}), in terms of the single grain
diffusion propagator $D_0(\Omega_n).$ Diagrams (b) define the
renormalization of the Coulomb vertex due to impurity scattering.
Diagrams (c) define the dynamically screened Coulomb interaction,
Eq.~(\ref{Vscreen}). The thin dashed lines represent the bare
Coulomb interaction while the thick lines denote the screened
Coulomb interaction.} \label{Coulomb}
\end{figure}

In this section we consider the suppression of the superconducting
transition temperature in granular metals due to combine effects
of Coulomb interaction and disorder (fermionic mechanism). The
Coulomb interaction in granular metals is screened by surrounding
electrons as in any metal.  The diagrams that describe the
screened effect are presented in Fig.~\ref{Coulomb}. The diagrams
(a) define the diffusion propagator in granular metals. As in the
case with the Cooperon, Eq.~(\ref{cooperon1}), we can consider the
single grain Diffusion in the zero dimensional approximation such
that $D_0^{-1}=\tau | \Omega_n | $ and for the diffusion
propagator we obtain the expression
\begin{equation}
D(\Omega _{n},\mathbf{q}) = \tau ^{-1}(|\Omega _{n}|+\varepsilon
_{\mathbf{q}}\delta )^{-1},  \label{diffuson}
\end{equation}
that coincides with the Cooperon propagator in
Eq.~(\ref{cooperon1}). The diagram (b) in Fig.~\ref{Coulomb}
describes the renormalization of the Coulomb vertex due to
impurities and the diagram (c) defines the screened Coulomb
interaction
\begin{equation}
\label{Vscreen} V(\Omega_n, {\bf q})=\left[ { { C({\bf q})} \over
{e^2}} + {{ 2 \varepsilon_{ \bf q}  }\over {|\Omega_n| +
\varepsilon_{\bf q}\delta }}  \right]^{-1},
\end{equation}
where $C(\bf q)$ is the Fourie transform of the capacitance matrix
which has the following asymptotic form at $q\ll a^{-1}$
\begin{equation}
C^{-1}(\mathbf{q})=\frac{2}{a^{d}}\,\left\{
\begin{array}{rl}
& \pi /q\hspace{1.4cm}d=2, \\
& 2\pi /q^{2}\hspace{1.1cm} d=3.
\end{array}
\right.   \label{capacitance}
\end{equation}

The correction to the critical temperature due to Coulomb
interaction before averaging over impurities is given by the
diagrams a-c in Fig~1. Averaging over the impurities leads to
rather complicated formulae. We will consider the contributions
from the diagrams a,b and c separately presenting the total
correction to the critical temperature due to Coulomb interaction
as
\begin{equation}
\left( \frac{\Delta T_{c}}{T_{c}}\right) _{f}=\langle X_{1}\rangle +\langle
X_{2}\rangle ,  \label{f}
\end{equation}
where the term $X_{1}$ represents the contribution of the diagrams a and b
in Fig.~1 while $X_{2}$ represents the contribution of the diagram c and $%
\langle ...\rangle $ means averaging over the disorder. The two
terms in the right hand side of Eq.~(\ref{f}) have a transparent
physical meaning: the term $X_{1}$ describes the renormalization
of the Cooperon due to Coulomb repulsion while the term $X_{2}$
describes the vertex renormalization. After the disorder averaging
the terms $\langle X_{1}\rangle $ and $\langle X_{2}\rangle $ are
represented in Figs.~\ref{fermionic1} and \ref{fermionic2},
respectively. The evaluation these diagrams is presented in
Appendix~B. The final expression for the correction to the
critical temperature due to the fermionic mechanism has the
following form
\begin{subequations}
\begin{widetext}
\begin{eqnarray}
\label{f1} \left(\frac{\Delta T_c}{T_c}\right)_f &=& -
4T\sum_{{\bf q}, \Omega_n > 0} V(\Omega_n,{\bf q}) \left[
F(\Omega_n)\, {{  \varepsilon_{\bf q} \delta}\over {(\Omega_n+
\varepsilon_{\bf q}\delta)^2\Omega_n}} +
 {1\over {4\pi T}} \,
{{\varepsilon_{\bf q} \delta}\over {\Omega_n^2 - (\varepsilon_{\bf
q}\delta)^2 } }\, \psi^\prime\left(\frac{1}{2} +
\frac{\Omega_n}{2\pi T}\right)
\right. \nonumber \\
&+& \left. 2F(\Omega_n)\, K(\Omega_n, {\bf q})\,
{{(\varepsilon_{\bf q}\delta)^2 }\over {[\Omega_n^2 -
(\varepsilon_{\bf q}\delta)^2]^2}} \, \left( \psi[\,1/2 +
(\Omega_n + \varepsilon_{\bf q}\delta)/4\pi T\,]- \psi[\,1/2 +
\Omega_n/2\pi T\, ] \right)  \right],
\end{eqnarray}
\end{widetext}
where we introduced the notation
\label{bosonic1}
\begin{equation}
F(\Omega _{n})=\psi (1/2+\Omega _{n}/2\pi T)-\psi (1/2),
\end{equation}
and the propagator $K(\Omega_n, {\bf q})$ was defined in
Eq.~(\ref{K1}). The summation in Eq.~(\ref{f1}) is going over the
quasi-momentum $\mathbf{q}$ and the bosonic Matsubara frequencies
$\Omega _{n}.$ The propagator of the screened electron-electron
interaction, $V(\Omega _{n},\mathbf{q})$, Eq.~(\ref{Vscreen}), in
the limit when the charging energy, $E_c$, is much larger than the
average mean level spacing, $\delta$, can we written as
\begin{equation}
V(\Omega _{n},\mathbf{q})={\frac{{2E_{C}(\mathbf{q})(|\Omega
_{n}|+\varepsilon _{\mathbf{q}}\delta )}}{{4\varepsilon
_{\mathbf{q}}E_{C}(\mathbf{q})+|\Omega _{n}|}}}.  \label{V11}
\end{equation}
\end{subequations}
where $E_{C}(\mathbf{q}) = e^{2}/2C(\mathbf{q})$ is the charging
energy.

\begin{figure}[t]
\includegraphics[width=.95\linewidth]{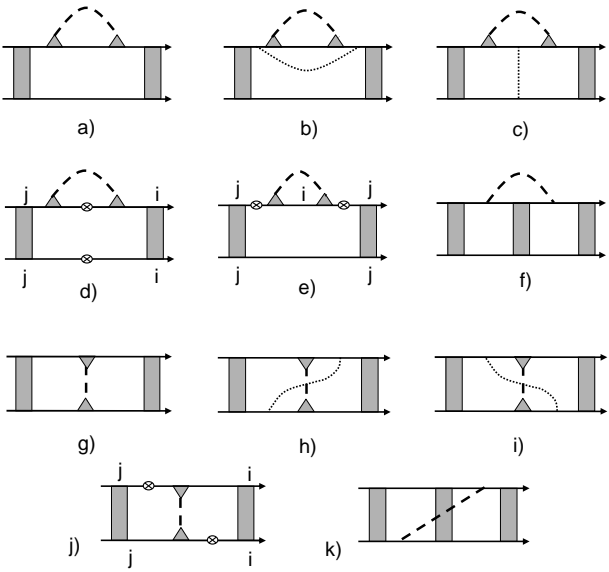}
\caption{Diagrams obtained from diagrams a and b in Fig.~1 after
disorder averaging. The solid lines denote the propagator of
electrons, the dashed lines describe screened Coulomb interaction
and the dashed-dot lines describe the elastic interaction of
electrons with impurities. The shaded rectangle and triangle
denote the renormalized Cooperon, see Eq.~(\ref{cooperon1}), and
impurity vertex of granular metals respectively. The indices $i$
and $j$ stand for the grain numbers. The tunneling vertices are
described by the circles.} \label{fermionic1}
\end{figure}

An important feature of Eq.~(\ref{f1}) is that the expression in
the big square brackets in the right hand side vanishes at ${\bf
q} \to 0$ for any frequency $\Omega_n.$ This is due to the fact
that the potential $V(\Omega_n, {\bf q} = 0)$ represents a pure
gauge and thus it should not contribute to the thermodynamic
quantities such as the superconducting critical temperature. As a
consequence one can see that the contribution of the frequencies
$\Omega_n$ that belong to the interval
\begin{equation}
\varepsilon_{\bf q} \delta < \Omega_n < \varepsilon_{\bf q} E_C
({\bf q}) \label{interval}
\end{equation}
have a small contribution to the critical temperature correction
given by Eq.~(\ref{f1}), \cite{Finkelstain_review}. For this
reason we do not expect the Coulomb energy $E_C$ to appear in the
final result. At the same time the frequencies $\Omega_n$ that
belong to the interval (\ref{interval}) are fully responsible for
the logarithmic renormalization of the integranular conductance
$g_{\scriptscriptstyle T}(T) = g_{\scriptscriptstyle T} - (1/ 2\pi
d ) \, \ln(g_TE_{\scriptscriptstyle C}/T )$, where the Coulomb
energy explicitly appears in the result.

The expression in the r.h.s. of Eq.~(\ref{f1}) is quite
complicated; we cannot derive a simple result at any arbitrary
temperature.  Further on we will  consider only the limiting cases
$T > \Gamma $ and $T < \Gamma$ where the calculations are
considerably simplified.

If the temperature $T$ is sufficiently small, $T\ll \Gamma $, the
summation over the Matsubara frequencies can be replaced with the
integral. One can easily see that the singularities at $\Omega =
\varepsilon_{\bf q} \delta$ in the second and third terms in
Eq.~(\ref{f1}) cancel each other. Their appearance, in fact,  is a
pure artifact of the representation of the result in terms of the
$\psi-$functions. With logarithmic accuracy one can leave only the
first term in Eq.~(\ref{f1}); this results in
\begin{equation}
 \left(\frac{\Delta T_c}{T_c}\right)_f =-{1\over {2 \pi }}
 \sum_{\bf q} {1\over {\varepsilon_{\bf q} }}\,
 \ln^2 {{ \varepsilon_{\bf q} \delta}\over { T }}, \;\;\;\;
 \varepsilon_{\bf q} \delta \gg T.   \label{log-ac}
\end{equation}
In the $3d$ case one can neglect the $q-$ dependence under the
logarithm in Eq.~(\ref{log-ac}), and the summation over
quasimomentum leads to the logarithmically accurate final
result~(\ref{mainresult_fermion}). In two dimensions, the main
contribution in summation over quasimomentum in Eq.~(\ref{log-ac})
comes from the low momenta $ q\ll 1/a$ where the energy
$\varepsilon_{\bf{q}}$ can be written as $
\varepsilon_{\bf{q}}=g_T {\bf q}^2 a^2$, and the granular system
becomes equivalent to a homogenously disordered one. Summation
over $q$ with the logarithmic accuracy leads to the final result
~(\ref{mainresult_fermion}) for 2D case. No wonder that the
result, Eq.~(\ref{mainresult_fermion}) in 2D agrees with the known
result for disorder metals
Refs.~\onlinecite{Ovchinnikov73,Fukuyama81,Finkelstein87,Finkelstain_review}.
Equation~(\ref{mainresult_fermion}) in the 2D case has a universal
form and is expressed in terms of the tunneling conductance
$g_{\scriptscriptstyle T}$ and the effective relaxation time
$\left( g_{\scriptscriptstyle T}\delta \right)^{-1}$. For the
homogenously disordered samples the latter time should be replaced
by the elastic scattering time $\tau $.

\begin{figure}[b]
\includegraphics[width=.95\linewidth]{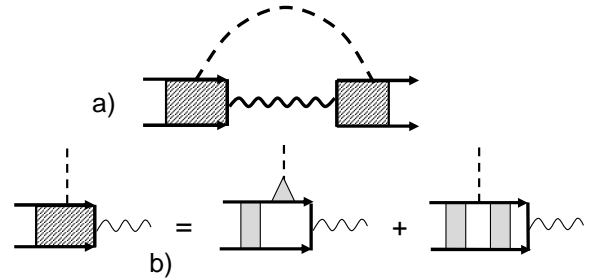}
\caption{Diagrams describing vertex renormalization obtained from
the diagram c) in Fig.~1 after averaging over the disorder. All
notations are the same as in Figs.~2 and 4.} \label{fermionic2}
\end{figure}

In the opposite limit, $T\gg \Gamma $ the quantity $\varepsilon
_{\bf q}\delta $ can be neglected with respect to the Matsubara
frequencies $\Omega _{n}$ and the result is drastically different
from the one given by Eq.~(\ref{mainresult_fermion}). The
potential $V(\Omega_n, {\bf q})$ in this limit takes the form
$V(\Omega_n, {\bf q}) = |\Omega_n| / 2 \varepsilon_{\bf q}$ such
that $\varepsilon_{\bf q}$ cancels in the main approximation.
Summation over Matsubara frequencies then leads to the correction
Eq.~(\ref{totalTc}).

One thus can see from the above that in the limit $T\gg \Gamma$
the fermionic mechanism of the suppression of the
superconductivity is no longer efficient. This can be seen rather
easily in another way using the phase approach of
Ref.~\onlinecite{Efetov02}. Following these works one decouples
the Coulomb interaction, Eq.~(\ref{Hc}) by integration over a
phase $\phi$
\begin{widetext}
\begin{eqnarray}
\exp \left( -\frac{e^{2}}{2} \int d\tau \sum_{ij} {n}_i(\tau)
C_{ij}^{-1} {n}_j(\tau) \right) &=& \int \exp \left(
-i\sum_{i}\int n_i(\tau) \dot{\phi}_i \left( \tau \right) d\tau -
\sum_{ij}\int d\tau \dot{\phi}_i \left( \tau \right)
\frac{C_{ij}}{2e^2}\dot{\phi}_j \left( \tau \right) \right) D\phi,
\label{c1}
\end{eqnarray}
\end{widetext}
where $n_i(\tau) =  \int \psi ^{\ast }\left(
\mathbf{r}_{\mathbf{i}},\tau \right) \psi \left( \mathbf{r}_i,\tau
\right) d\mathbf{r}_i$ is the electron number in the i-th grain
and $\psi$ are fermionic fields. With this decoupling the variable
$\dot{ \phi}$ plays a role of an additional chemical potential. In
the limit $T\gg \Gamma $ (and only in this limit) the phase $\phi
$ can be gauged out via the replacement
\begin{equation}
\psi \left( \mathbf{r}_{i},\tau \right) \rightarrow \psi \left(
\mathbf{r} _{i},\tau \right) \exp \left[ -i\phi _{i}\left( \tau
\right) \right]. \label{c3}
\end{equation}
Substituting Eq.~(\ref{c3}) into Eq.~(\ref{hamiltonian1} -
\ref{Ht}) (or to be more precise, into the corresponding
Lagrangian in the functional integral representation) we
immediately see that the phase $\phi $ enters the tunneling term,
Eq.~(\ref{Ht}) only. However, this term is not important in the
limit $T\gg \Gamma$ and we conclude that the long range part of
the Coulomb interaction leading to charging of the grains is
completely removed. Therefore, the effect of the Coulomb
interaction on the superconducting transition temperature must be
small and this is seen from Eq.~(\ref{totalTc}). This conclusion
matches well the fact that the upper limit in the logarithms in
Eq.~(\ref{mainresult_fermion}) is just $\Gamma $ and at
temperatures exceeding this energy the logarithms should
disappear. At low temperatures $T < \Gamma $ the phase description
does not apply and  the non-trivial result,
Eq.~(\ref{mainresult_fermion}), appears. This result is of the
pure quantum origin, and interference effects are very important.
In the limit of high temperatures $T > \Gamma $ the interference
effects are suppressed, that is why the fermionic mechanism of the
suppression of the superconductivity is no longer efficient.

\section{Discussion}

We have described the suppression of the superconducting
transition temperature due to (i) fluctuations of the order
parameter (bosonic mechanism) and (ii) Coulomb repulsion
(fermionic mechanism) in granular
metallic systems at large tunneling conductance between the grains, $g_{%
\scriptscriptstyle T}\gg 1$. We have calculated the correction to
the transition temperature for $3d$
granular samples and films. We have demonstrated that at temperature $T>g_{%
\scriptscriptstyle T}\delta $ the suppression of superconductivity
in granular metals is determined by the bosonic mechanism while at
low temperatures, $T<g_{\scriptscriptstyle T}\delta $, the
suppression of superconductivity is dominated by the fermionic
mechanism.

The bosonic mechanism has a classical origin. The fermionic
mechanism is of the quantum origin and is relevant at low
temperature $T<\Gamma $ where quantum interference effects are
pronounced. In the opposite limit $T>\Gamma $ the coherence is
lost and the fermionic mechanism of suppression of the
conductivity is no longer efficient. Thus the classical bosonic
mechanism is more important at high temperatures, whereas the
quantum fermionic one is efficient at low temperatures.

The results of the theoretical study presented in this paper can
be checked experimentally. Apparently, the best way to do this is
to study the dependence of the superconducting transition
temperature $T_{c}$ of granular metals as a function of the
tunneling conductance $g_{\scriptscriptstyle T}$. Practically this
can be done by studying several granular samples with different
tunneling conductances (different oxidation coating). The
experimental curves for $T_{c}$ suppression should have a
different slope at high $T_{c}>g_{\scriptscriptstyle T}\delta$ and
low $T_{c}<g_{\scriptscriptstyle T}\delta $ critical temperatures
due to the fact that the suppression of the superconductivity is
given by the two different mechanisms. Moreover, the difference
between the dependence of the critical temperature on the
tunneling conductance $T_{c}(g_{\scriptscriptstyle T})$ in
granular and homogeneously disordered systems can give an
information on the morphology of the sample, {\it i.e.} answer the
question whether the samples are homogeneously disordered or
granular.

Another interesting consequence of our results in
Eqs.~(\ref{sumall}) is the following: since at low critical
temperatures $T_{c}<g_{\scriptscriptstyle T}\delta $ the
suppression of superconductivity in granular metals is given by
the fermionic mechanism and upon the substitution
$g_{\scriptscriptstyle T}\delta \rightarrow \tau ^{-1}$ it
coincides with the proper result for homogeneously disordered
samples,~\cite{Ovchinnikov73} one can generalize the
renormalization group result by Finkelstein~\cite{Finkelstein87}
for the $T_{c}$ suppression. The latter result obtained for
homogeneously disordered films can be applied to the case of the
granular superconductors upon the proper substitution for the
diffusion coefficient $D=g_{\scriptscriptstyle T}\delta a^{2}$,
where $a$ is the size of a single grain.

\begin{acknowledgments}
We thank Igor Aleiner for useful and stimulating discussions.  We
thank M.~Feigelman for pointing us out on the importance of the
bosonic mechanism. This work was supported by the U.~S.~Department
of Energy, Office of Science through contract No.~W-31-109-ENG-38,
by the SFB-Transregio~12 of German Research Society, and by DIP
and GIF Projects of German-Israeli Cooperation.
\end{acknowledgments}

\appendix

\section{ Evaluation of diagrams for the bosonic mechanism }

\begin{figure}[t]
\resizebox{.45\textwidth}{!}{\includegraphics{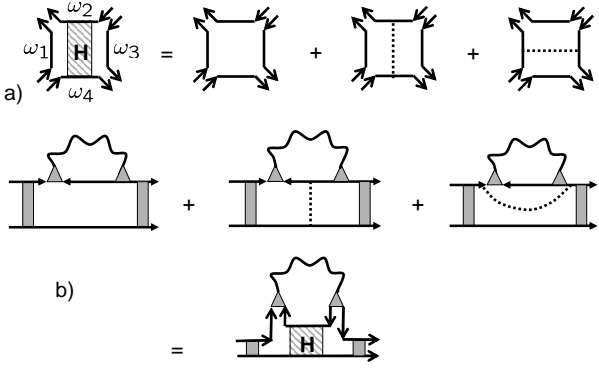}}
\vspace{0.6cm} \caption{Diagrams $a)$ describe "bosonic" Hikami
box, Eq.~(\ref{Hikami}). Using Hikami box the sum of three
diagrams a), b) and c) in Fig.~\ref{BososnicFig}   can be
conveniently represented as a single diagram shown in Fig.~6b. All
notations are the same as in Fig. 3. } \label{Hikami_Boson}
\end{figure}

In this appendix we evaluate diagrams for bosonic mechanism
presented in Fig.~\ref{BososnicFig} beginning with consideration
of the diagrams a), b) and c). These three diagrams can be
conveniently combined in a single diagram introducing the Hikami
box~\cite{Hikami81} as it is shown  in Fig.~\ref{Hikami_Boson}.
For zero dimensional grain (all characteristic energies are much
less than the Thouless energy) Hikami box is given by the
following expression
\begin{equation}
\label{Hikami}
H(\varepsilon_{n_1},\varepsilon_{n_2},\varepsilon_{n_3},\varepsilon_{n_4})
= 2\pi \tau^2 (|\varepsilon_{n_1}| + |\varepsilon_{n_2}| +
|\varepsilon_{n_3}| + |\varepsilon_{n_4}|),
\end{equation}
where $\varepsilon_{n_i}$ are fermionic Matsubara frequencies.
Using Eq.~(\ref{Hikami}) and evaluating the diagram shown in
Fig.~6b we obtain the following result for the sum of these three
diagrams
\begin{equation}
\label{A2}
 - \pi T^2 \delta
\sum\limits_{\varepsilon_n(\varepsilon_n - \Omega_n)
> 0} \frac{3|\varepsilon_n| + |\Omega_n - \varepsilon_n|}{\varepsilon_n^2( |2\varepsilon_n - \Omega_n| +
\varepsilon_{\bf q}\delta )^2} K(\Omega_n, \mathbf{q}),
\end{equation}
where $K(\Omega_n, \mathbf{q})$ is the propagator of
superconducting fluctuations defined in Eq.~(\ref{K1}).

The sum of the diagrams d) and e) in Fig.~\ref{BososnicFig}  is
given by
\begin{equation}
\label{A3} - \pi T^2 \delta
\sum\limits_{\varepsilon_n(\varepsilon_n - \Omega_n)
> 0} \frac{K(\Omega_n, \mathbf{q})}{\varepsilon_n^2
( |2\varepsilon_n - \Omega_n| + \varepsilon_{\bf q}\delta  )^2} \,
\varepsilon_{\bf q}\delta.
\end{equation}

The above expression for diagrams d) and e) in
Fig.~\ref{BososnicFig} in fact includes an additional contribution
coming from the diagrams a), b) and c) in Fig.~\ref{BososnicFig}
that was not included in Eq.~(\ref{A2}). This additional
contribution appears due to the fact that the single electron
Green function self-energy has a correction resulting from the
renormalization of the Green function self-energy due to
intergranular tunneling, see Eq.~(\ref{self}) and Fig.~4. This
self energy correction is negligible in the diffusive limit
($\tau\to 0$), nevertheless it gives a finite contribution to the
sum of the diagrams a) - c) because each of these diagrams diverge
in the diffusive limit as $1/\tau$ while their sum remains finite
due to cancellation of the leading orders in $1/\tau$. The
additional finite term appears because the diagrams b) and c) have
impurity lines that are determined by the bare mean free time
$\tau_0$ while in all other places the mean free time appears
through the Green function self energy that contains the
renormalized $\tau.$ This additional contribution could have  been
written as an extra constant term $8 \pi d g_T \delta$  in the
Hikami box. We, however, find it is natural to "redirect" this
term to the diagrams d) and e) since these diagrams are also
proportional to the tunnelling conductance $g_T.$

Finally, for the diagram f) in Fig.~\ref{BososnicFig} we obtain
\begin{equation}
\label{A4}  \pi T^2 \delta
\sum\limits_{\varepsilon_n(\varepsilon_n - \Omega_n) < 0}
\frac{K(\Omega_n, \mathbf{q})}{\varepsilon_n
^2}\frac{1}{|\Omega_n| + \varepsilon_{\bf q}\delta }.
\end{equation}
Adding all the contributions given by Eqs.~(\ref{A2})-(\ref{A4})
we obtain the correction to the superconducting transition
temperature due to bosonic mechanism presented in Eq.~(\ref{K}).

\section{Evaluation of diagrams for the fermionic mechanism}

In this appendix we evaluate diagrams for fermionic mechanism
presented in Figs.~\ref{fermionic1} and \ref{fermionic2} . We
begin our analysis with the evaluation of the contribution
$\langle X_1 \rangle$ in the right hand side of Eq.~(\ref{f}) that
can be written as a sum of two terms
\begin{subequations}
\label{bosonic1}
\begin{equation}  \label{X}
\langle X_1 \rangle = \langle X_1^a \rangle + \langle X_1^b \rangle,
\end{equation}
where $\langle X_1^a \rangle$ and $\langle X_1^b \rangle$
represent the contributions of the diagrams a-f and g-j in
Fig.~\ref{fermionic1} respectively. The sum of the diagrams a), b)
and c) can be presented as a single diagram with the help of the
Hikami box shown in Fig.~\ref{Hikami_Fermion}a exactly as in the
case of the bosonic diagrams considered in Appendix A.  The
corresponding Hikami box shown in Fig.~\ref{Hikami_Fermion}a
differs from the "bosonic" Hikami box only by the arrow directions
and is given by the same Eq.~\ref{Hikami}. The sum of the diagrams
a), b) and c) in Fig.~\ref{fermionic1} thus results in the
following expression
\begin{figure}[b]
\resizebox{.45\textwidth}{!}{\includegraphics{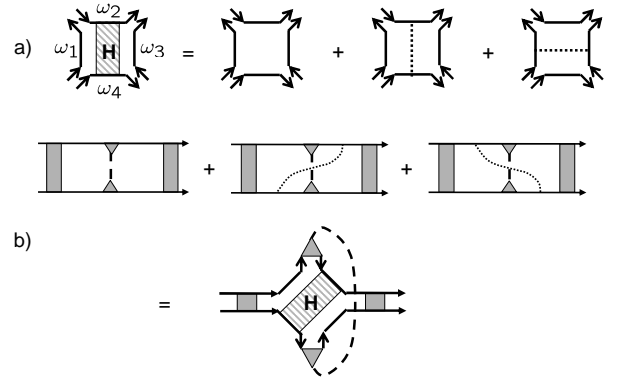}}
\vspace{0.6cm} \caption{ Diagrams a) describe "fermionic" Hikami
box. Using Hikami box the sum of three diagrams g), h) and i) in
Fig.~\ref{fermionic1} can be conveniently represented as a single
diagram b). All notations are the same as in Fig.~\ref{fermionic1}
.} \label{Hikami_Fermion}
\end{figure}
\begin{equation}
\label{B1} - 2\pi T^2 \sum\limits_{\varepsilon_n(\varepsilon_n -
\Omega_n)<0} \frac{\left(|\varepsilon_n| + |
\Omega_n|/2\right)}{\varepsilon_n^2\,(|\Omega_n| +
\varepsilon_{\bf q}\delta)^2 } V(\Omega_n, {\bf q}),
\end{equation}
where summation is going over the quasi-momentum, $\mathbf{q}$, fermionic, $%
\varepsilon_n$ and bosonic, $\Omega_n$ Matsubara frequencies. The
propagator of the screened electron-electron interaction,
$V(\Omega_n ,\mathbf{q})$ in Eq.~(\ref{B1}) was defined in
Eq.~(\ref{V11}).

The sum of the diagrams d) and e) in Fig.~\ref{fermionic1} results
in the following contribution
\begin{equation}
-\pi T^2 \sum\limits_{\varepsilon_n(\varepsilon_n - \Omega_n)<0}
\frac{V(\Omega_n, {\bf q})}{\varepsilon_n^2\,(|\Omega_n| +
\varepsilon_{\bf q}\delta)^2 } \, \varepsilon_{\bf q}\delta,
\end{equation}
while the diagram f) is given by
\begin{equation}
\label{B2} \pi T^2 \sum\limits_{\varepsilon_n(\varepsilon_n -
\Omega_n)
> 0} \frac{ V(\Omega_n, {\bf q})
}{\varepsilon_n^2\,(|2\varepsilon_n- \Omega_n| + \varepsilon_{\bf
q}\delta )}.
\end{equation}
Summing up all contributions in Eqs.~(\ref{B1})-(\ref{B2}) we
arrive to the following expression that represents the
contribution of diagrams a)-f) in Fig.~\ref{fermionic1}
\begin{widetext}
\begin{equation}
\langle X_1^a \rangle = - \pi T^2 \sum\limits_{{\bf q}} \left[
\sum\limits_{\varepsilon_n(\varepsilon_n - \Omega_n)<0} \left(
\frac{2V(\Omega_n, {\bf q})}{|\varepsilon_n|\,(|\Omega_n| +
\varepsilon_{\bf q}\delta)^2 } + \frac{ V(\Omega_n, {\bf
q})}{\varepsilon_n^2(|\Omega_n| + \varepsilon_{\bf q}\delta
)}\right) - \sum\limits_{\varepsilon_n(\varepsilon_n -\Omega_n)>
0}\frac{V(\Omega_n, {\bf q})}{\varepsilon_n^2(|2 \varepsilon_n-
\Omega_n| + \varepsilon_{\bf q}\delta)} \right]. \hspace{1cm}
\label{X1a}
\end{equation}
\end{widetext}

Now we turn to the evaluation of the diagrams g-j in
Fig.~\ref{fermionic1} that are represented by the term $\langle
X_1^b \rangle$ in Eq.~(\ref{X}). Calculation of the sum of the
diagrams g), h) and i) again can be reduced to the evaluation of a
single diagram shown in Fig.~7b resulting in
\begin{equation}
\label{B1f}
-\pi T^2 \sum\limits_{\varepsilon_n(\varepsilon_n
-\Omega_n)<0} \frac{|\Omega_n| \, \,V(\Omega_n, {\bf
q})}{(|\Omega_n| + \varepsilon_{\bf q}\delta)^2 |\varepsilon_n|
|\varepsilon_n -\Omega_n| }.
\end{equation}
The diagram j) in Fig.~\ref{fermionic1} is given by the following
expression
\begin{equation}
-\pi T^2 \sum\limits_{\varepsilon_n(\varepsilon_n -\Omega_n)<0}
\frac{\varepsilon_{\bf q} \delta \, \, \, V(\Omega_n, {\bf
q})}{(|\Omega_n| + \varepsilon_{\bf q}\delta)^2 |\varepsilon_n|
|\varepsilon_n -\Omega_n| },
\end{equation}
while the diagram k) results in
\begin{equation}
\label{B1h} -\pi T^2 \sum\limits_{\varepsilon_n(\varepsilon_n
-\Omega_n)>0} \frac{V(\Omega_n, {\bf
q})}{(|2\varepsilon_n-\Omega_n| + \varepsilon_{\bf q}\delta)
|\varepsilon_n| |\varepsilon_n -\Omega_n| }.
\end{equation}
Summing up the above contributions, Eqs.~(\ref{B1f}) -
(\ref{B1h}), we obtain the expression representing the sum of the
diagrams g) - k) in Fig.~\ref{fermionic1}
\begin{widetext}
\begin{equation}
\langle X_1^b \rangle = - \pi T^2 \sum\limits_{{\bf q}}\left[
\sum\limits_{\varepsilon_n(\varepsilon_n -\Omega_n)<0}
\frac{V(\Omega_n, {\bf q})}{|\varepsilon_n||\varepsilon_n
-\Omega_n|(|\Omega_n| + \varepsilon_{\bf q}\delta)} +
\sum\limits_{\varepsilon_n(\varepsilon_n -\Omega_n)>0}
\frac{V(\Omega_n, {\bf q})}{ |\varepsilon_n||\varepsilon_n
-\Omega_n|(|2\varepsilon_n - \Omega_n| + \varepsilon_{\bf q}\delta
)} \right] \label{X1b}.
\end{equation}
\end{widetext}

Now we turn to the evaluation of the vertex renormalization which
is given by the term $\langle X_2 \rangle$ in the right hand side
of Eq.~(\ref{f}). The corresponding diagrams are shown in
Fig.~\ref{fermionic2}. Averaging over impurities results in the
renormalization of the effective interaction vertex between the
Coulomb and Cooper pair propagators. The renormalized vertex
$\Gamma(\Omega_n)$ is given by the sum of two diagrams shown in
Fig.~\ref{fermionic2}b that lead to the following expression
\begin{widetext}
\begin{equation}
\label{Gamma} \Gamma(\Omega_n) = {T\over{\Omega_n +
\varepsilon_{\bf q} \delta}} \sum_{0 < \varepsilon_n < \Omega_n}
{1\over \varepsilon_n} \, + \, T \sum_{\varepsilon_n > 0} \left(
{1\over \varepsilon_n} - {1\over {\varepsilon_n +
\Omega_n}}\right) \, {1\over {2\varepsilon_n + \Omega_n +
\varepsilon_{\bf q}\delta }}.
\end{equation}
\end{widetext}
Using Eq.~(\ref{Gamma}) the resulting expression for the term $\langle X_2
\rangle$ in Eq.~(\ref{f}) can be written as
\end{subequations}
\begin{equation}
\langle X_2 \rangle = 8\pi^2 T \sum_{\mathbf{q}, \Omega_n > 0} V(\Omega_n,
\mathbf{q}) \, K(\Omega_n, \mathbf{q}) \Gamma^2(\Omega_n),  \label{X2}
\end{equation}
where $K(\Omega_n, \mathbf{q})$ is the propagator of
superconducting fluctuations defined in Eq.~(\ref{K1}). Expressing
summations over the fermionic frequencies in
Eqs.~(\ref{X1a},~\ref{X1b}) in terms of the di-gamma functions
after some rearrangements of different terms in Eqs.~(\ref
{X1a},~\ref{X1b}) we obtain Eq.~(\ref{f1}) for the suppression of
superconducting transition temperature due to fermionic mechanism.


\vspace{-0.4cm}\vspace{-0.2cm}

\begin{thebibliography}{99}

\bibitem{Valles} K.~L.~Ekinci and J.~M.~Valles, Jr., Phys. Rev. Lett.
\textbf{\textbf{82}}, 1518 (1999).

\bibitem{experiment} A.~Gerber, A.~Milner, G.~Deutscher, M.~Karpovsky,
and A.~Gladkikh, Phys. Rev. Lett.~\textbf{78}, 4277 (1997).

\bibitem{Jaeger} H.~M.~Jaeger, D.~B.~Haviland, A.~M.~Goldman, and B.~G.~Orr,
Phys. Rev. B \textbf{\textbf{34}}, 4920 (1986).

\bibitem{Simon} R.~W.~Simon, B.~J.~Dalrymple, D.~Van~Vechten,
W.~W.~Fuller, and S.~A.~Wolf, Phys. Rev. B~\textbf{36}, 1962
(1987).

\bibitem{Beloborodov99} I.~S.~Beloborodov and K.~B.~Efetov,
Phys. Rev. Lett.~\textbf{82}, 3332 (1999).

\bibitem{Efetov02} K.~B.~Efetov and A.~Tschersich,
Europhys. Lett.~\textbf{59}, 114, (2002); Phys. Rev. B
\textbf{67}, 174205 (2003).

\bibitem{Anderson59} P.~W.~Anderson, J. Phys. Chem. Solid {1}, 26 (1959).

\bibitem{Ovchinnikov73} Yu.~N.~Ovchinnikov,
Zh. Eksp. Teor. Fiz. \textbf{64}, 719 (1973) [Sov. Phys.
JETP~\textbf{37}, 366 (1973)].

\bibitem{Fukuyama81} S.~Maekawa and H.~Fukuyama,
J. Phys. Soc. Jpn.~\textbf{51}, 1380 (1982); S.~Maekawa,
H.~Ebisawa and H.~Fukuyama, J. Phys. Soc. Jpn.~\textbf{52}, 1352
(1983).

\bibitem{Finkelstein87} A.~M.~Finkelstein, Pis'ma Zh. Eksp. Teor. Fiz.
\textbf{45}, 37 (1987) [Sov. Phys. JETP Lett.~\textbf{45}, 46 (1987)];
Phisica B \textbf{197}, 636 (1994).

\bibitem{Ishida98} H. Ishida and R. Ikeda, J.~Phys.~Soc.~Jpn. \textbf{67}, 983 (1998);
R.~A.~Smith, B.~S.~Handy and V.~Ambegaokar, Phys.~Rev.~B
\textbf{61}, 6352 (2000).

\bibitem{Larkin99} A.~I.~Larkin, Ann. Phys.~\textbf{8}, 785
(1999).

\bibitem{Efetov80} K.~B.~Efetov, Zh.~Eksp.~Teor.~Fiz.~\textbf{78}, 2017
(1980) [Sov. Phys. JETP~\textbf{51}, 1015 (1980)].

\bibitem{Fisher90} M.~P.~A.~Fisher, Phys. Rev. Lett.~\textbf{65}, 923 (1990).

\bibitem{Efetov} I.~S.~Beloborodov, K.~B.~Efetov, A.~Altland and
F.~W.~J.~Hekking, Phys. Rev. B~\textbf{63}, 115109 (2001).

\bibitem{Lopatin03} I.~S.~Beloborodov, K.~B.~Efetov, A.~V.~Lopatin and
V.~M.~Vinokur, Phys. Rev. Lett.~\textbf{91}, 246801 (2003).

\bibitem{Altshuler} B.~L.~Altshuler and A.~G.~Aronov, in \textit{\
Electron-Electron Interaction in Disordered Systems}, ed. by A.~L.~Efros and
M.~Pollak, North-Holland, Amsterdam (1985).

\bibitem{Universal} I.~S.~Beloborodov, A.~V.~Lopatin, and V.~M.~Vinokur,
Phys.~Rev.~B,~\textbf{70}, 205120 (2004).

\bibitem{AGD} A.~A.~Abrikosov, L.~P.~Gorkov and I.~E.~Dzyaloshinskii,
\textit{Methods of Quantum Field Theory in Statistical Physics}
(Prentice-Hall, Englewood Cliffs, NJ, 1963).

\bibitem{Finkelstain_review} A.~M.~Finkelstein, \textit{Electron liquid in
Disordered Conductors}, edited by I.~M.~Khalatnikov, Soviet Scientific
Reviews Vol.~\textbf{\ 14} (Harwood, London, 1990).

\bibitem{BLV_2004} I.~S.~Beloborodov, A.~V.~Lopatin, and V.~M.~Vinokur,
Phys. Rev. Lett. \textbf{92}, 207002 (2004).

\bibitem{Aleiner} We thank Igor Aleiner for pointing this out.

\bibitem{Hikami81} S.~Hikami, Phys. Rev. B \textbf{24}, 2671 (1981).

\end{thebibliography}
\end{document}